\documentclass[prd,showpacs,nofootinbib,
12pt]{revtex4}

\usepackage{amssymb, amsfonts, amsmath}
\usepackage{graphicx}

\newcommand{\lb}{\label}
\newcommand{\be}{\begin{equation}}
\newcommand{\ee}{\end{equation}}
\newcommand{\e}{{\rm e}}
\newcommand{\const}{{\rm const}}
\usepackage{color}

\begin{document}

\title{Higher-$n$ triangular dilatonic black holes}
\author{Anton Zadora} \email{as.zadora@physics.msu.ru}
\affiliation{Department of Theoretical Physics, Faculty of Physics, Moscow State University, 119899, Moscow, Russia}

\author{Dmitri V. Gal'tsov} \email{galtsov@phys.msu.ru}
\affiliation{Department of Theoretical Physics, Faculty of Physics, Moscow State University, 119899, Moscow, Russia}
\affiliation{Kazan Federal University, 420008 Kazan, Russia}

\author{Chiang-Mei Chen} \email{cmchen@phy.ncu.edu.tw}
\affiliation{Department of Physics, National Central University, Chungli 32001, Taiwan}

\begin{abstract}
Dilaton gravity with the form fields is known to possess dyon solutions with two horizons for the discrete (``triangular'') values of the dilaton coupling constant $a = \sqrt{n (n + 1)/2}$. From this sequence only $n = 1,\, 2$ members were known analytically so far. We present two new $n = 3,\, 5$ triangular solutions for the theory with different dilaton couplings $a,\, b$ in electric and magnetic sectors in which case the quantization condition reads $a b = n (n + 1)/2$. These are derived via the Toda chains for $B_2$ and $G_2$ Lie algebras. Solutions are found in the closed form in general $D$ space-time dimensions. They satisfy the entropy product rules and have negative binding energy in the extremal case.
\end{abstract}


\maketitle

\section{Introduction}
Einstein-Maxwell-dilaton (EMD) theory in four dimensions may originate from different supersymmetric higher-dimensional theories with various values of the dilaton coupling constant $a$ in the Maxwell term $\e^{-2 a \phi} F^2$ in the Lagrangian. For $a = 0$ the dilaton decouples and the theory reduces to Einstein-Maxwell (EM) system, which is the bosonic part of $N = 2,\, D = 4$ supergravity. The value $a = 1$ corresponds to $N = 4,\, D = 4$ supergravity~\cite{Gibbons:1982ih, Gibbons:1987ps}. The value $a = \sqrt{3}$, corresponds to dimensionally reduced $D = 5$  gravity~\cite{Gibbons:1985ac,Rasheed:1995zv}, which have different supersymmetric extensions. In all these cases analytical solutions are known for static black holes possessing both electric and magnetic charges (dyons) with two horizons between which the dilaton exhibits $n$ oscillations. Such solutions have non-singular extremal limits with AdS$_2 \times S^2$ horizons contrary the one-charged dilatonic black hole which have singular horizons in the extremal limit~\cite{Gibbons:1982ih, Gibbons:1987ps, Garfinkle:1990qj} for generic coupling constant.

As it was first noticed by Poletti, Twamley and Wiltshire~\cite{Poletti:1995yq}, the values $a = 1,\, \sqrt{3}$ are just the two lowest members $n = 1,\, 2$ of the ``triangular'' sequence of dilaton couplings
\begin{equation} \label{an}
a_n = \sqrt{n (n + 1)/2},
\end{equation}
for which dyons (known for higher $n$ only numerically) exhibit similar behavior of the dilaton. Analytically this triangle quantization rule was rederived in~\cite{Galtsov:2014wxl} as condition of regularity of the dilaton in the case of coinciding horizons, i.e. for extremal dyons. The same rule was recently shown to arise from the linearized dilaton equation on the Reissner-Nordtstr\"om dyonic background~\cite{Davydov:2017zbs} as condition of existence of dilaton bound states. However, exact higher-$n$ dyonic solutions were not known analytically. Here we give solutions for $n = 3,\, 5$ in the theory where the dilaton couples to  electric and magnetic sectors with different coupling constants $a,\, b$, in which case the quantization rule generalizes to $a b = n (n + 1)/2$. We also give generalization to arbitrary dimensions. These new solutions are derived via Toda chains for $B_2$ and $G_2$ algebras. The solutions satisfy the entropy product rule:  the product of the entropies of the internal and internal horizons are functions of their charges only \cite{Cvetic:1996xz,Cvetic:2013eda}. This property is considered to be an indication on possibility of statistical interpretation of the entropy. Another interesting feature is that the extremal dyons have negative binding energy like the $SL(n,R)$ multi-scalar Toda black holes \cite{Lu:2013uia}.

Previous work on Toda dilatonic black holes includes among other~\cite{Lu:1996hh,  Ivashchuk:1999xp, Ivashchuk:2000yc, Lu:1997rd}. Our results   have some overlap with the recent preprints~\cite{Abishev:2017hmd, Davydov:2017zbs}.

\section{Toda black holes}
\subsection{Setup}
Consider Einstein-Maxwell-dilaton system in $D$ dimensions
\begin{equation} \label{theoryAction}
S = \int d^D x \sqrt{-g} \left( R - \frac{1}{2} \partial_\mu \phi \partial^\mu \phi - \frac{1}{2 (D-2)!} \e^{a \phi} F_{[D-2]}^2 - \frac{1}{4} \e^{b \phi} F_{[2]}^2 \right),
\end{equation}
with two different dilaton coupling constants $a, b$ for the magnetic $(D-2)$-form $F_{[D-2]}$ and the electric $2$-form $F_{[2]}$. Assume the static spherically symmetric ansatz for the metric
\begin{equation}
ds^2 = -\e^{2B} dt^2 + \e^{2A} dr^2 + \e^{2C} d\Omega_{D-2}^2,
\end{equation}
and the following solution of the equations of motion for the form fields:
\begin{equation} \label{forms}
F_{[D-2]} = P \, \mathrm{vol}(\Sigma_{D-2}), \qquad  F^{rt} = Q \e^{-b \phi} \e^{- A - B - (D - 2) C}.
\end{equation}
All the unknown functions $A,\, B,\, C,\, \phi$ depend on a single variable $r$. One convenient gauge choice is $A(r) = - B(r)$. Denoting then $\e^{C(r)} = R(r)$, from (\ref{theoryAction}) one can derive the following four equations for three functions $B,\, R,\, \phi$ (one of which following from the other three by  the Bianchi identity):
\begin{eqnarray} \label{system3}
\left( R^{D-2} (\e^{2 B})' \right)' &=& \frac{D-3}{(D-2) R^{D-2}} \left( P^2 \e^{a \phi} + Q^2 \e^{-b \phi} \right),
\nonumber \\
\left( \e^{2 B} (R^{D-2})' \right)' &=& -\frac{1}{2 R^{D-2}} \left( P^2 \e^{a \phi} + Q^2 \e^{-b \phi} \right) + (D-2) (D-3) R^{D-4},
\nonumber \\
R'' + \frac{1}{2 (D-2)} R \phi'^2 &=& 0,
\nonumber \\
\phi'' + \phi' \left( \e^{2 B} R^{D-2} \right)' \e^{-2 B} R^{-(D-2)} &=& \frac{1}{2} \e^{-2 B} \left( a P^2 \e^{a \phi} - b Q^2 \e^{-b \phi} \right) R^{-2(D-2)}.
\end{eqnarray}

\subsection{Triangular quantization}
The dilaton coupling quantization can be obtained analytically for extremal dyons as condition of regularity of the dilaton at the degenerate event horizon $r = r_0$, where the expansion of $\e^{2B}$ in $x \equiv (r - r_0)/r_0$ starts with $x^2$:
\begin{eqnarray} \label{expansions}
R &=& R_h + \rho_1 x + \rho_2 x^2 + O(x^3),
\nonumber \\
\e^{2B} &=& \nu_2 x^2 + O(x^3),
\nonumber \\
\phi &=& \phi_h + \mu_n x^n + O(x^{n+1}), \quad n \in \mathbb{Z}.
\end{eqnarray}
The leading power index $n$ of the dilaton is assumed to be integer to ensure analyticity of the solution. The second and the third equations in~(\ref{system3}) yield the following expressions for $R_h$ and $\rho_2$
\begin{eqnarray}
R_h^{2(D-3)} &=& \frac{P^2 \e^{a \phi_h} + Q^2 \e^{-b \phi_h}}{(D-3) (D-2)},
\nonumber\\
\rho_2 &=& - \frac{R_h \mu_n^2}{4 (D-2)} \delta_{n,1}.
\end{eqnarray}
The other equations lead to the following relations
\begin{eqnarray} \label{relations}
a P^2 \e^{a \phi_h} - b Q^2 \e^{-b \phi_h} &=& 0,
\nonumber \\
2 \nu_2 n (n+1) &=& \frac{r_0^2}{R_h^{2(D-2)}} \left( a^2 P^2 \e^{a \phi_h} + b^2 Q^2 \e^{-b \phi_h} \right),
\nonumber \\
2 \nu_2 &=& \frac{(D-3) r_0^2}{(D-2) R_h^{2 (D-2)}} \left( P^2 \e^{a \phi_h} + Q^2 \e^{-b \phi_h} \right).
\end{eqnarray}
Using the first equation to simplify the ratio of the second and the third ones,  one finally obtains the necessary condition for the product of coupling constants:
\begin{equation} \label{quantization}
a b = n (n + 1) \frac{D-3}{D-2}.
\end{equation}
Redefining them as
\begin{equation} \label{redefinition}
\left( a,\, b \right)=\lambda( \tilde{a},\, \tilde{b} ) , \qquad \lambda^2=  \frac{2 (D-3)}{D-2},
\end{equation}
we get
\begin{equation}
\tilde{a} \tilde{b} = \frac{n(n+1)}{2},
\end{equation}
which for $a = b$ coincides with~(\ref{an}).

\subsection{Toda representation}
Passing to another gauge:
\begin{equation}
\e^{2C} = \e^{2A} f r^2, \qquad f = 1 - \frac{2 \mu}{r^{D-3}},
\end{equation}
one obtains
$
\e^{2A} = \e^{-\frac{2B}{D-3}} f^{-\frac{D-4}{D-3}},
$
so the metric reduces to
\begin{equation} \label{ansatz}
ds^2 = -\e^{2B} dt^2 + \e^{-\frac{2B}{D-3}} f^{\frac{1}{D-3}} \left( f^{-1} dr^2 + r^2 d\Omega_{D-2}^2 \right).
\end{equation}
The function $f$, often called the blackening factor, has one zero corresponding to the black-hole horizon and located at $r=r_+,\,r_+^{D-3} = 2 \mu$. The inner horizon resides at $r = 0$, and for the extremality parameter $\mu=0$ the event horizon coincides with the inner horizon.
Switching to the new radial variable $\rho$,
\begin{equation} \label{rho}
\e^{- 2 (D-3) \mu \rho} = 1 - \frac{2 \mu}{r^{D-3}}, \qquad \partial_\rho = f r^{D-2} \partial_r,
\end{equation}
and denoting the derivative $\partial_\rho$ by dot, one arrives at the following lagrangian:
\begin{equation} \label{Leff}
\mathcal{L} = \dot{B}^2 + \frac{\lambda^2}4 \dot{\phi}^2 + \frac{\lambda^2}4 \e^{2 B} \left( P^2 \e^{a \phi} + Q^2 \e^{-b \phi} \right),
\end{equation}
which belongs to the Toda type. This Lagrangian has a discrete $S$-duality symmetry:
\begin{equation} \label{Sd}
Q \leftrightarrow P, \qquad a \leftrightarrow -b.
\end{equation}
With this in mind, we can restrict to magnetically dominated solutions $P > Q$, obtaining the electrically dominated $Q > P$ via this duality.

The general two-component Toda system for the variables $\chi_j,\, j = 1,\, 2$
\begin{equation} \label{TodaLBC}
\mathcal{L} = \frac{1}{2} \sum_{i, j = 1}^2 B_{i j} \ \dot{\chi_i} \dot{\chi_j} + \sum_{i = 1}^2 g_i^2 \e^{\sum_{j=1}^2 C_{i j} \chi_j}
\end{equation}
is known to be integrable if the matrix $C_{i j}$ is the Cartan matrix of some Lie algebra and the matrix $K := B^{-1} C^T$ is diagonal~\cite{Perelomov}. For irreducible rank two algebras one has three options: $A_2,\, B_2$ and $G_2$, since $C_2 \simeq B_2$, and $D_2 \simeq A_1 \times A_1$ is reducible. In all these cases the Cartan matrix can be represented as
\begin{equation} \label{Cartan}
C = \begin{pmatrix} 2 & -1 \\ C_{21} & 2 \end{pmatrix},
\end{equation}
where $C_{21} = -1$ for $A_2$, $C_{21} = -2$ for $B_2$ or $C_{21} = -3$ for $G_2$. To put our system into this form one makes the substitution
\begin{eqnarray} \label{2Bphi}
2B &=& \frac{1}{\tilde{a} + \tilde{b}} \left[ (2 \tilde{b} + \tilde{a} C_{21}) \chi_1 + (2 \tilde{a} - \tilde{b}) \chi_2 \right],
\nonumber\\
\lambda\phi &=& \frac{1}{\tilde{a} + \tilde{b}}  \left[ (2 - C_{21}) \chi_1 - 3 \chi_2 \right],
\end{eqnarray}
arriving at~(\ref{TodaLBC}) with
\begin{equation}
g_1^2 = \frac{\lambda^2}4  P^2 , \qquad g_2^2 =\frac{\lambda^2}4  Q^2,
\end{equation}
and the matrix $B$ given by
\begin{eqnarray} \label{Bmatrix}
&& B_{11} = \frac{4 (1 + C_{21} + \tilde{b}^2) + C_{21} \left[ -4 + C_{21} (1 + \tilde{a})^2 \right]}{4 (\tilde{a} + \tilde{b})^2}, \qquad B_{22} = \frac{9 + (\tilde{b} - 2 \tilde{a})^2}{4 (\tilde{a} + \tilde{b})^2},
\nonumber \\
&& B_{12} = B_{21} = \frac{3 (-2 + C_{21}) + (2 \tilde{a} - \tilde{b}) (2 \tilde{b} + C_{21} \tilde{a})}{4 (\tilde{a} + \tilde{b})^2}.
\end{eqnarray}

Imposing the conditions of diagonality on the matrix $K := B^{-1} C^{T}$ we find:
\begin{eqnarray} \label{Cartanab}
&& C_{21} = - \frac{1 + \tilde{b}^2}{1 + \tilde{a}^2}, \qquad \tilde{a} = \frac{3 + \tilde{b}^2}{2 \tilde{b}},
\nonumber \\
&& K = B^{-1} C^{T} = \frac2{B_{22}} \begin{bmatrix} - C_{21}^{-1} & 0 \\ 0 & 1 \end{bmatrix}.
\end{eqnarray}
Substituting the values of $C_{21}$ for Lie algebras $A_2, B_2, G_2$ one gets:
\begin{eqnarray} \label{A2C2G2couplings}
A_2: && C_{21} = -1, \quad \tilde{a} = \tilde{b} = \pm \sqrt{3},
\nonumber\\
B_2: && C_{21} = -2, \quad \tilde{a} = \pm 2, \quad \tilde{b} = \pm 3,
\nonumber\\
G_2: && C_{21} = -3, \quad \tilde{a} = \pm {5}/\sqrt{3}, \quad \tilde{b} = \pm 3 \sqrt{3}.
\end{eqnarray}
Obviously one can interchange $\tilde{a},\, \tilde{b}$ due to the symmetry. From~(\ref{A2C2G2couplings}) wee see that these values satisfy the quantization condition~(\ref{quantization}) for $n = 2,\, 3,\, 5$ respectively. The equations of motion for $\chi_1,\, \chi_2$ are
\begin{eqnarray} \label{chiequations}
\ddot\chi_1 &=& \kappa_1 g_1^2 \ \e^{2 \chi_1 - \chi_2},
\nonumber\\
\ddot\chi_2 &=& \kappa_2 g_2^2 \ \e^{C_{21} \chi_1 + 2 \chi_2},
\end{eqnarray}
where we denoted
\begin{equation}
\kappa_i \equiv K_{ii}, \qquad \kappa_1 = \tilde a^2 + 1, \quad \kappa_2 = \tilde b^2 + 1.
\end{equation}

To proceed further we make the following decomposition:
\begin{equation} \label{chidecomposition}
\e^{\chi_1} = H_1^{-1} f^{\gamma_1}, \qquad \e^{\chi_2} = H_2^{-1} f^{\gamma_2}.
\end{equation}
Here $H_1,\, H_2$ are new unknown functions and $\gamma_1,\, \gamma_2$ are rational powers. For non-extreme solutions we have to impose the conditions $g_{tt} \propto f^1$ (non-degeneracy of the horizon) and $\e^\phi \propto f^0$ (regularity of the  horizon) leading to the following expressions for $\gamma_1,\, \gamma_2$:
\begin{equation}
\gamma_1 = \frac{3}{4 + C_{21}}, \qquad \gamma_2 = \frac{2 - C_{21}}{4 + C_{21}}.
\end{equation}
Note that the parameters $\gamma = (\gamma_1,\, \gamma_2)$ are nothing but the components of the dual Weyl vector for given Lie algebra with the Cartan matrix~(\ref{Cartan}). They may be represented in a explicitly symmetric form with respect to the discrete S-duality~(\ref{Sd}):
\begin{equation}
 \gamma_1 = \frac{\tilde{b}(1 + \tilde{a}^2)}{2 (\tilde{a} + \tilde{b})}, \qquad \gamma_2 = \frac{\tilde{a}(1 + \tilde{b}^2)}{2 (\tilde{a} + \tilde{b})}.
\end{equation}
In terms of the new variables the metric of the general solution reads:
\begin{eqnarray} \label{intervalgeneral}
ds^2 &=& - H_1^{-h_1} H_2^{-h_2} f dt^2 + H_1^\frac{h_1}{D-3} H_2^\frac{h_2}{D-3} \left( f^{-1} dr^2 + r^2 d\Omega_{D-2}^2 \right),
\nonumber \\
\e^{\lambda\phi} &=& H_1^{\tilde{a} h_1}\, H_2^{-\tilde{b} h_2},
\end{eqnarray}
where
$h_1 = 2/ \kappa_1,\; h_2 = 2/ \kappa_2$, so one has the following relation between the parameters introduced
\begin{equation}\label{hh}
    h_1 \gamma_1 + h_2 \gamma_2 = 1.
\end{equation}
The extremal solutions correspond to the limit $\mu \to 0$.

The unknown functions $H_1,\, H_2$ obey the following system of equations written in terms of the original radial coordinate $r$:
\begin{equation} \label{mainequationsD}
r^{D-2} \left[ r^{D-2} f (\ln H_i)' \right]' + \kappa_i g_i^2 \prod_{j=1}^2 H_j^{-C_{ij}} = 0.
\end{equation}
Assuming $H_i \to 1$ as $r \to \infty$, we look for the polynomial solutions of the form
\begin{eqnarray} \label{Hrepresentation}
H_1 = 1 + \frac{P_1}{r^{D-3}} + \frac{P_2}{r^{2(D-3)}} + \cdots + \frac{P_p}{r^{p(D-3)}}, && p = 2 \gamma_1,
\nonumber\\
H_2 = 1 + \frac{Q_1}{r^{D-3}} + \frac{Q_2}{r^{2(D-3)}} + \cdots + \frac{Q_q}{r^{q(D-3)}}, && q = 2 \gamma_2.
\end{eqnarray}
where the number of terms depends on a given algebra.  The polynomials $H_1,\, H_2$ are interchanged by S-duality  as follows:
\begin{equation}\label{polynomialsSymmetry}
P_i \leftrightarrow Q_i , \qquad p \leftrightarrow q, \qquad \kappa_1 g_1^2 \leftrightarrow \kappa_2 g_2^2.
\end{equation}

Consider behavior of the metric as $r\to 0$. The dominant terms in the polynomials $H_1,\,H_2$ are $P_p/\varrho^p$ and $Q_q/\varrho^q$ respectively, where $\varrho=r^{(D-3)}$. The line element in (\ref{intervalgeneral}) for $\mu=0$ with account for (\ref{hh}) reduces to
\begin{equation}\label{r0}
ds^2=-c \varrho^2   dt^2 + c^{1/(3-D)}\left( (D-3)^{-2}\frac{d\varrho^2}{\varrho^2}+d\Omega_{D-2}^2\right)
\end{equation}
with $c=P_p^{-h_1}Q_q^{-h_2}$, which is obviously $AdS_2\times S^{D-2}$. This is the geometry of the extremal horizon as expected. The spatial section of the hypersurface $r=0$ therefore is the sphere $S^{D-2}$. For $\mu\neq 0$ the interval in the vicinity of $r=0$ is
\begin{equation}\label{r0}
ds^2=2\mu c \varrho dt^2 + c^{1/(3-D)}\left(  -2\mu (D-3)^{-2}\frac{d\varrho^2}{\varrho^3}+d\Omega_{D-2}^2\right),
\end{equation}
so $\varrho>0 $ is $T$-region. The $D-2$ sphere $\varrho=0$ is the internal horizon, which is again non-singular.

In the non-extreme case it is  easier to work in another representation. Specifically, for $\mu \neq 0$ one can use the following map~\cite{Ivashchuk:2013jja}:
\begin{equation} \label{map}
H_i(r) = \frac{\mathcal{H}_i (z)}{\mathcal{H}_i(1)} ,
\end{equation}
where $z \equiv f = 1 - 2\mu/r^{D-3}$. Then the functions $\mathcal{H}_i$ obey the equations
\begin{equation} \label{EqHi}
\frac{d}{d z} \left[ z \frac{d}{d z} \ln\mathcal{H}_i(z) \right] = \mathcal{P}_{i} \prod_{j = 1}^2 \mathcal{H}_j^{-C_{ij}},
\end{equation}
provided the following relation holds for $\mathcal{P}_i$:
\begin{equation} \label{constraint}
\left[ 2 \mu (D-3) \right]^2 \mathcal{P}_{i} \prod_{j = 1}^2 \mathcal{H}_j(1)^{-C_{ij}} = - \kappa_i g_i^2.
\end{equation}
We will use this condition to establish relations between charges $P, Q$ and coefficients $p_1, q_1$.
The functions $\mathcal{H}_i$ may also be represented as polynomials of $z$ of degrees $p,\, q$:
\begin{equation}
\mathcal{H}_1(z) = 1 +  2 \gamma_1 \sum_{j=i}^{p} p_i z^i, \qquad \mathcal{H}_2(z) = 1 +  2 \gamma_2 \sum_{i = 1}^{q} q_i z^i,
\end{equation}
In the Eqs.~(\ref{EqHi}, \ref{constraint}) $\mathcal{P}_{i}$ may be set arbitrary. We fix  $\mathcal{P}_i = ( 2 \gamma_1 p_1,\, 2 \gamma_2 q_1)$ to ensure the simplest form of the polynomials $\mathcal{H}_i$. Note again that the map~(\ref{map}) is valid only as long as $\mu \ne 0$, so the extreme case has to be considered separately.

\subsection{Physical parameters}
The mass of the solutions in our gauge can be extracted from the asymptotic $g_{tt}=1-2M/{r^{D-3}}+\ldots$ as $r\to\infty$. It depends on the extremality parameter $\mu$ and the first two coefficients of the polynomials $P_1,\,Q_1$ as follows:
\be  \label{mass}
2M =   2 \mu + h_1 P_1 + h_2 Q_1 .
\ee
Similarly, from the asymptotic of the dilaton function $\e^{\lambda\phi} = 1 +  {2 \Sigma}/{r^{D-3}}+\ldots $
 we find:
\begin{equation}
    2 \Sigma = \tilde{a} h_1 P_1 - \tilde{b} h_2 Q_1.
\end{equation}

From the Lagrangian (\ref{Leff}) one can derive the hamiltonian which is constant by virtue of the equations of motion:
\begin{equation} \label{Heff}
\mathcal{H} = \dot{B}^2 + \frac{\lambda^2}4 \dot{\phi}^2 -\frac{\lambda^2}4 \e^{2 B} \left( P^2 \e^{a \phi} + Q^2 \e^{-b \phi} \right),
\end{equation}
where dot denotes derivative with respect to the variable $\rho$ introduced in (\ref{rho}). We can equate $\mathcal{H}_+=\mathcal{H}(\rho\to\infty)$, corresponding to the horizon $r=r_+$, and $\mathcal{H}_\infty=\mathcal{H}(\rho=0)$, corresponding to $r\to\infty)$. Using the polynomial form of $\e^B,\, \e^{\lambda\phi}$ and transforming to $\rho$, one finds:
\begin{equation}
    \mathcal{H}_\infty = (D-3)^2\left( M^2 + \Sigma^2\right)   - \lambda^2(P^2 + Q^2).
\end{equation}
On the horizon one has: $ \e^{2B} = 0, \; \dot{\phi} = 0, \; \dot{B} = - \mu (D-3)$ entailing $\mathcal{H}_+=\mu^2(D-3)^2$.
Thus from $\mathcal{H}_+=\mathcal{H}_\infty$ one obtains the following relation between the physical parameters of the solution
\begin{equation} \label{chargemassrelation}
(2 M)^2 + (2 \Sigma)^2 - (2 \mu)^2= \frac{2(P^2 + Q^2)}{(D-2)(D-3)},
\end{equation}
which reduces to  the familiar no-force condition~\cite{Poletti:1995yq,Galtsov:2014wxl} for $\mu=0$ (note different choices of units: $16\pi G=1$ here and  $4\pi G=1$ in~\cite{Poletti:1995yq,Galtsov:2014wxl}).
We will use the equation~(\ref{chargemassrelation}) to check  our results presented below.

The Bekenstein-Hawking entropy associated with the outer $r=r_+$ and inner $r=0$ horizons is computed as the quarter of their areas (as we have seen, the inner horizon is not a point but a $(D-2)$-sphere):
\be \lb{Entr}
S_\pm  =  \frac{\pi^\frac{D-1}2}{2 \Gamma\left( \frac{D-1}2 \right)} r_\pm^2 \left[ H_1(r_\pm) \right]^\frac{h_1}{D-3} \, \left[ H_2(r_\pm) \right]^\frac{h_2}{D-3},
\ee
where $r_+^{D-3} = 2 \mu$ and $r_- = 0$.
Actually, $S_-$ is expressed through the last polynomial coefficients $P_p$ and $Q_q$:
\begin{equation} \label{residualentropy}
S_- =  \frac{\pi^\frac{D-1}2}{2 \Gamma\left( \frac{D-1}2 \right)} P_p^\frac{h_1}{D-3} Q_q^\frac{h_2}{D-3}.
\end{equation}
The Hawking temperature is computed in terms of the surface gravity of the event horizon:
\be
T_H  =   \frac{D-3}{4 \pi r_+} \left[ H_1(r_+) \right]^{- h_1/\lambda^2} \, \left[ H_2(r_+) \right]^{- h_2/\lambda^2} .
\ee
The relation between the physical charges and coefficients $p_1,\, q_1$ or, equivalently, $P_1,\, Q_1$ may be obtained with help of the Eq.~(\ref{constraint}).

In the extreme limit $\mu \to 0$ the event horizon merges with the surface $r=0$ while the polynomials $H_1,\,H_2$ are such that $g_{tt}\sim r^{2(D-3)}$. Then the temperature vanishes, but the entropy is still finite and is given again by (\ref{residualentropy}). In this case the relevant coefficients
are proportional to the powers of the charges: $P_p \sim P^p$, $Q_q \sim Q^q$, therefore
\be S_{{\rm ext}} \sim \lvert P \rvert ^{p\frac{ h_1}{D-3}} \lvert Q \rvert^{q\frac{h_2}{D-3}}.
\ee
For $n=2$ the following entropy rule is known to hold~\cite{Lu:2013uia}:
\be
S_{-}S_{+} = S_{{\rm ext}}^2,
\ee
so the intriguing question is whether this is also true for $B_2$ and $G_2$ solutions. We will see that this is indeed the case.  This property is regarded as indication on the microscopic origin of the entropies involved~\cite{Cvetic:1996xz,Cvetic:2013eda}.

\section{Solutions in $D = 4$}
In this section we present solutions for $A_2,\, B_2,\, G_2$ algebras in $D = 4$ with non-trivial dilaton profile. The first case is well-known \cite{Gibbons:1985ac}, we include it here just as illustration of our representation. The dilaton oscillates between the two horizons, exhibiting $n$ zeroes. This behavior is in agreement with the results of~\cite{Poletti:1995yq} obtained numerically for $a = b$.    Note that  due to our parametrization of the metric~(\ref{ansatz}) the same formulas will be valid for any  $D$, see the last subsection.  Since the solutions were obtained with fixed values of the coupling constants $a, b$ in accord with~(\ref{A2C2G2couplings}), the S-duality under the transformation~(\ref{polynomialsSymmetry}) is not manifest explicitly.

It is worth noting that our equations also have trivial $\phi = \const$  solutions for charges obeying the relation~$a P^2 = b Q^2$, which are  Reissner-Nordstrom dyons.

\subsection{$A_2 \simeq sl(3, R)$}
In this well-known case we have $p = q = 2$ and the polynomials $\mathcal{H}_i$ have the following structure:
\begin{eqnarray}
\mathcal{H}_1 &=& 1 + 2 p_1 z + p_1 q_1 z^2,
\nonumber \\
\mathcal{H}_2 &=& 1 + 2 q_1 z + p_1 q_1 z^2.
\end{eqnarray}
One can also obtain expressions of $p_1,\, q_1$ via the coefficients $P_1,\, Q_1$ for $H_1(r),\, H_2(r)$:
\begin{equation}
p_1 = -\frac{P_1 (Q_1 + 2 \mu)}{(P_1 + 2 \mu) (Q_1 + 4 \mu)}, \qquad q_1 = -\frac{Q_1 (P_1 + 2 \mu)}{(P_1 + 4 \mu) (Q_1 + 2 \mu)},
\end{equation}
so the charges by virtue of the Eqs.~(\ref{constraint}) are
\begin{equation}
P^2=-2 r_+^2 p_1 \mathcal{H}_1(1)^{-2} \mathcal{H}_2(1) , \qquad  Q^2=-2r_+^2 q_1 \mathcal{H}_1(1) \mathcal{H}_2(1)^{-2}.
\end{equation}

\begin{figure}[ht]
\centering
\includegraphics[scale=1.0]{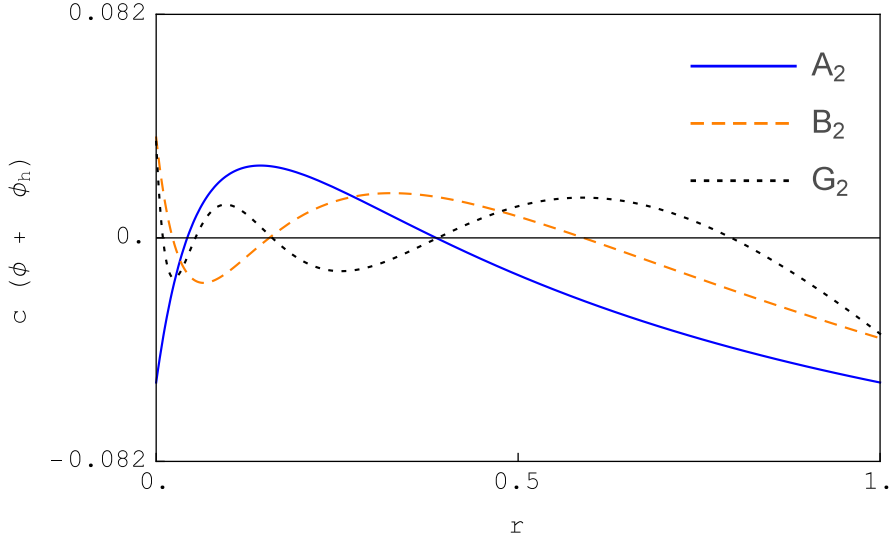}
\caption{ Behavior of the shifted dilaton field between the horizons for non-extreme $A_2$, $B_2$ and $G_2$ solutions for $r_+ = 1.0,\, P = 1.0,\, Q = \xi P,\, \xi = 0.5$. For better visualization, the dilaton is multiplied by $c=1/3$  for $A_2$,  and by  $c=10$  for $G_2$ solutions.}
\label{Phi}
\end{figure}

\begin{figure}[ht]
\centering
\includegraphics[scale=1.0]{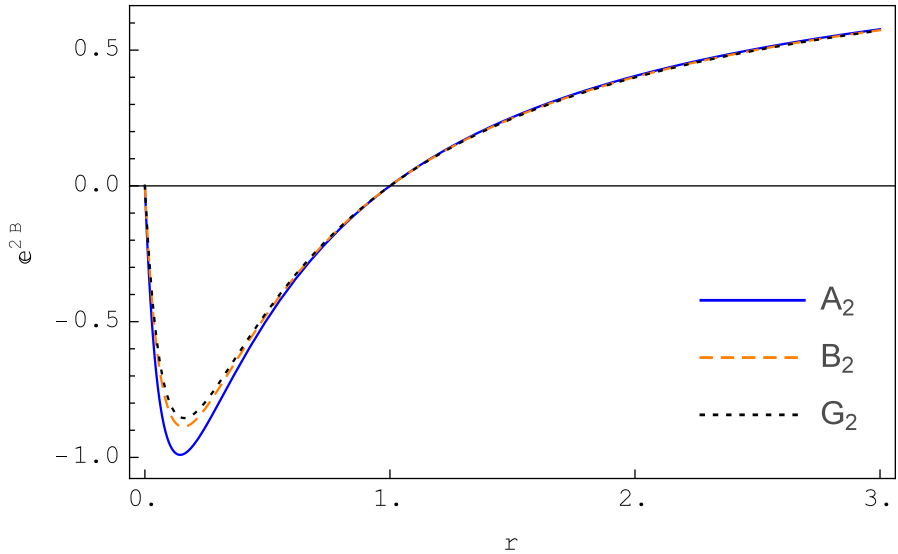}
\caption{ The function $\e^{2B}$ for non-extreme $A_2$, $B_2$ and $G_2$ solutions for $r_+ = 1.0,\, Q = \xi P,\, \xi = 0.5$.}
\label{Exp2B}
\end{figure}

In the extreme case ($\mu = 0$) we work with polynomials $H_i$ instead of $\mathcal{H}_i$. We can directly express coefficients $P_i,\, Q_i$ in polynomials $H_i$ through values of charges:
\begin{eqnarray}
&& P_1 = \sqrt{1 + \xi^{2/3}} \, P, \qquad Q_1 = \xi^{2/3} \sqrt{1 + \xi^{2/3}} \, P,
\nonumber \\
&& P_2 = \frac{1}{2} \xi^{2/3} \, P^2, \qquad Q_2 = \frac{1}{2} \xi^{4/3} \, P^2, \qquad Q = \xi P.
\end{eqnarray}
From here one finds the following expansion of the dilaton field $\phi$ near the horizon in the extremal case:
\begin{equation}
\phi_{{\rm ext}} = \frac{\sqrt3}{2} \ln\left(\frac{P_1}{Q_1} \right) + \sqrt3 \left(\frac{1}{P_1^2} - \frac{1}{Q_1^2} \right) r^2 + O(r^3).
\end{equation}
Typical behavior of the shifted dilaton $\phi + \phi_h$ and $\e^{2B}$ for the extremal $A_2,\, B_2$ and $G_2$ solutions is shown in Figs.~\ref{Phi},~\ref{Exp2B}, where $\phi_h$ is the horizon value in the extremal limit (note that it is different from $\phi(r_+)$ in the non-extreme case)

The mass  and  the residual entropy of the extremal solutions are:
\begin{equation} \label{A2mass}
 M = \frac{1}{4} \left( P^{2/3} + Q^{2/3} \right)^{3/2},   \qquad S_{{\rm ext}} = \frac{\pi}{2} |P Q| .
\end{equation}
The product of entropy of inner horizon $S_{-}$ and entropy of outer horizon $S_{+}$ obeys the following relation
\begin{equation}
 S_{-} S_{+} = \pi^2 r_+^4 \frac{p_1 q_1}{\mathcal{H}_1(1)\mathcal{H}_2(1)} = S_{{\rm ext}}^2.
\end{equation}

\begin{figure}[ht]
\centering
\includegraphics[scale=1.0]{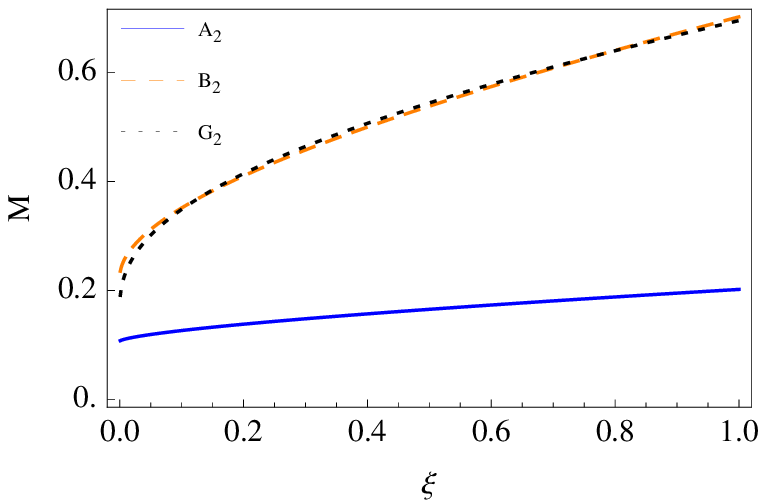}
\includegraphics[scale=1.0]{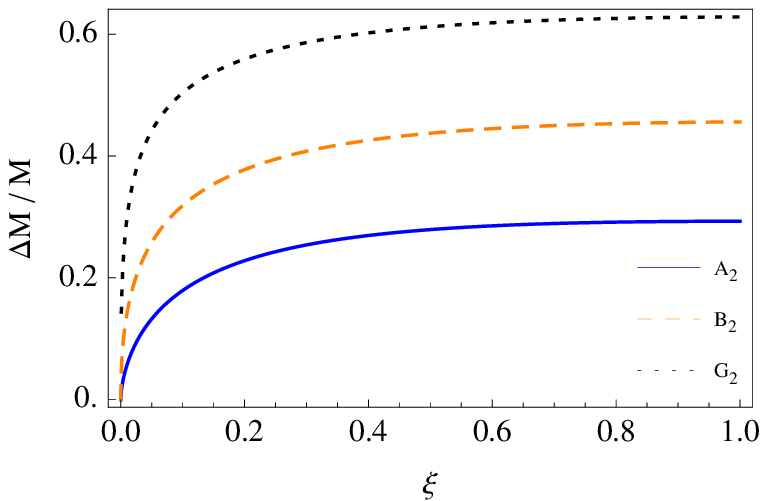}
\caption{The mass (left panel) and the normalized mass difference (right panel) of extremal magnetically dominated ($Q < P$) dyons as function of $\xi = Q/P\; (P = 1)$.}
\label{massesExtreme}
\end{figure}

From ~(\ref{A2mass}) it is easy to see that the solution has negative binding energy in a sense that the mass of the dyon is greater than the sum of masses of the singly charged electric and magnetic black holes with the same values of charges:
\begin{equation} \label{DeltaM}
\Delta M = M(Q, P) - M(Q, 0) - M(0, P),
\end{equation}
The  mass of dyons and the normalized mass difference~(\ref{DeltaM}) of extremal solutions are shown in Fig.~\ref{massesExtreme}. Black holes related to $SL(n, R),\, n \ge 3$ Toda chains were shown to have with negative binding energy (defined as $-\Delta M$) in~\cite{Lu:2013uia}. Such dyons could split into a pair of singly charged black holes.  For non-extremal solutions the sign of the binding energy depends on the charges.

\begin{figure}[ht]
\centering
\includegraphics[scale=1.0]{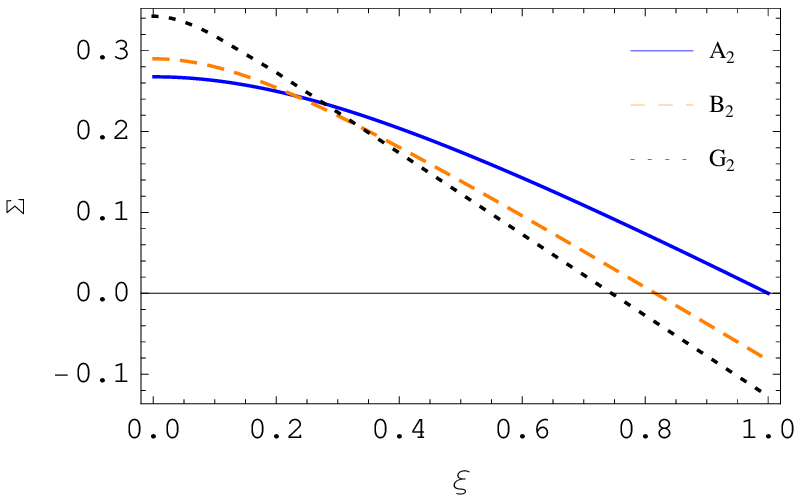}
\includegraphics[scale=1.0]{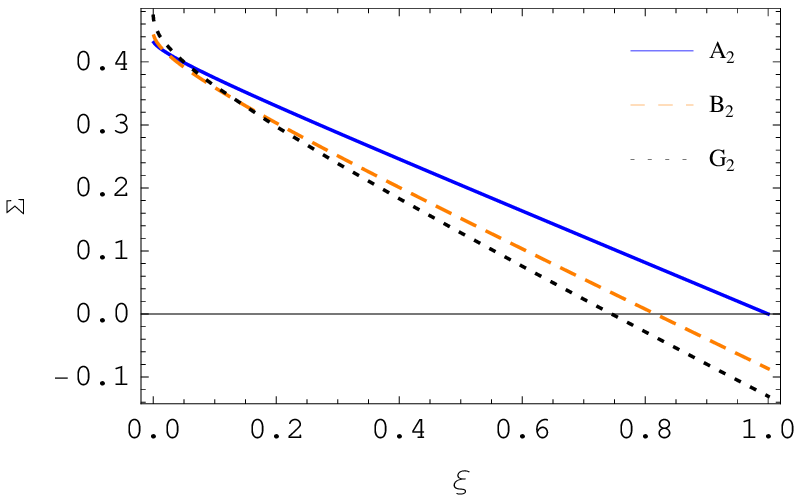}
\caption{Dilaton charge for extreme (left panel) and non-extreme (right panel) magnetically dominated ($Q < P$) solutions ($P = 1.0$ and  $r_+ = 1$ for non-extreme case).}
\label{scalarCharge}
\end{figure}

The dependence of the dilaton charge on parameter $\xi = Q/P$ in both extreme and non-extreme cases is shown in Fig.~\ref{scalarCharge}.

\subsection{$B_2 \simeq so(2, 3)$ }

Expressions for the polynomials $\mathcal{H}_i$ read
\begin{eqnarray}
\mathcal{H}_1 &=& 1 + 3 p_1 z + 3p_1 q_1 z^2 + p_1^2 q_1 z^3,
\nonumber \\
\mathcal{H}_2 &=& 1 + 4 q_1 z + 6 p_1 q_1 z^2 + 4 p_1^2 q_1 z^3 + p_1^2 q_1^2 z^4.
\end{eqnarray}
The relation between the pairs of the coefficients $p_1,\, q_1$ and $P_1,\, Q_1$ is
\begin{eqnarray}
p_1 &=& -\frac{(P_1^2 + 6 \mu P_1 + 6 \mu^2) (Q_1 + 4 \mu) - 2 \mu \Delta}{(P_1 + 2 \mu) (P_1 + 6 \mu) (Q_1 + 6 \mu)},
\nonumber \\
q_1 &=&  - \frac{(P_1^2 + 6 \mu P_1 + 10 \mu^2) (Q_1^2 + 8 \mu Q_1 + 8 \mu^2) + 16 \mu^4 - 2 \mu (Q_1 + 4 \mu) \Delta}{(P_1 + 4 \mu)^2 (Q_1 + 2 \mu) (Q_1 + 8 \mu)},
\nonumber \\
\Delta &:=&  \sqrt{P_1 (P_1 + 2 \mu) (P_1 + 4 \mu) (P_1 + 6 \mu) + (P_1 + 3 \mu)^2 (Q_1 + 4 \mu)^2},
\end{eqnarray}
and the constraint~(\ref{constraint}) gives
\begin{equation}
3 r_+^2 p_1 \mathcal{H}_1(1)^{-2} \mathcal{H}_2(1) = - \frac54 P^2, \qquad 4 r_+^2 q_1 \mathcal{H}_1(1)^2 \mathcal{H}_2(1)^{-2} = - \frac52 Q^2.
\end{equation}

In the extreme case we were able to express the coefficients $P_1,\, Q_1$ through the charges in terms of some parameter $\zeta$ instead of $\xi=Q/P$:
\begin{equation}
Q_1 = \frac{1 - \zeta^2}{2 \zeta} P_1, \qquad \zeta \le 1, \qquad \xi = \frac{(1 - \zeta)^2}{\sqrt{8 \zeta^3}}.
\end{equation}
We obtain:
\begin{eqnarray}
&& P_1 = \frac{1}{2} \left( \frac{5}{\zeta} \right)^{1/2} \, P, \qquad P_2 = \frac{1 - \zeta}{2} \, P_1^2, \qquad Q_2 = \frac{(1 - \zeta)^2}{2 \zeta} \, P_1^2,
\nonumber \\
&& P_3 = \frac{(1 - \zeta)^2}{12} \, P_1^3, \qquad Q_3 = \frac{(1 - \zeta)^3}{6 \zeta} \, P_1^3, \qquad Q_4 = \frac{(1 - \zeta)^4}{48 \zeta} \, P_1^4.
\end{eqnarray}
The near-horizon expansion of the dilaton in the extreme case is
\begin{equation}
\phi_{{\rm ext}} = \phi_h - \frac{4 \left[ (2 P_1^2 - Q_1^2) \sqrt{P_1^2 + Q_1^2} + (2 P_1^3 - Q_1^3) \right]}{P_1^3 Q_1^3} r^3 + O(r^4),
\end{equation}
with $\phi_h$ being constant (nonzero) value. This is exactly what we expected from~(\ref{expansions}).

The entropy of the extreme solution is given by the following formula:
\begin{equation}
  S_{{\rm ext}} = \frac{5 \pi}{4 \cdot 2^{2/5} 3^{3/5}} P^{6/5} Q^{4/5}.
\end{equation}
 The product of entropy of inner horizon $S_{-}$ and entropy of outer horizon $S_{+}$ obeys the following relation
\begin{equation}
S_{-} S_{+} = \pi^2 r_+^4 \frac{p_1^{6/5} q_1^{4/5}}{\mathcal{H}_1(1)^{4/5} \mathcal{H}_2(1)^{2/5}} = S_{{\rm ext}}^2.
\end{equation}

The numerical calculations confirms that this solution also has negative binding energy in the extreme case (see Fig.~\ref{massesExtreme}).

\subsection{$G_2$ solution}

The polynomials $\mathcal{H}_i$ are:
\begin{eqnarray}
\mathcal{H}_1 &=& 1 + 6 p_1 z + 15 p_1 q_1 z^2 + 20 p_1^2 q_1 z^3 + 15 p_1^3 q_1 z^4 + 6 p_1^3 q_1^2 z^5 + p_1^4 q_1^2 z^6,
\nonumber \\
\mathcal{H}_2 &=& 1 + 10 q_1 z + 45 p_1 q_1 z^2 + 120 p_1^2 q_1 z^3 + p_1^2 q_1 (135 p_1 + 75 q_1) z^4 + 252 p_1^3 q_1^2 z^5
\nonumber \\
&& + p_1^3 q_1^2 (75 p_1 + 135 q_1) z^6 + 120 p_1^4 q_1^3 z^7 + 45 p_1^5 q_1^3 z^8 + 10 p_1^6 q_1^3 z^9 + p_1^6 q_1^4 z^{10}.
\end{eqnarray}
The constraint~(\ref{constraint}) gives
\begin{equation}
6 r_+^2 p_1 \mathcal{H}_1(1)^{-2} \mathcal{H}_2(1) = - \frac73 P^2, \qquad 10 r_+^2 q_1 \mathcal{H}_1(1)^3 \mathcal{H}_2(1)^{-2} = - 7 Q^2.
\end{equation}
For $G_2$ algebra the expressions relating $p_1,\, q_1$ to $P_1,\, Q_1$ are rather lengthy, so we do not give them here.

 In the extreme  $G_2$-case we were not able to obtain short expressions using the auxiliary parameter $\xi$ (or $\zeta$), so we give here just the explicit relations between the coefficients $P_i,\, Q_i$:
\begin{eqnarray} \label{G2coeffs}
&& P_2 = \frac{1}{6} \left( -7 P^2 + 3 P_1^2 \right), \qquad Q_2 = \frac{1}{2} \left( -21 Q^2 + Q_1^2 \right),
\nonumber \\
&& P_3 = \frac{1}{18} \left( -7 P^2 Q_1 + 6 P_1 P_2 \right), \qquad Q_3 = \frac{1}{6} \left( -63 Q^2 P_1 + 2 Q_1 Q_2 \right),
\nonumber \\
&& P_4 = \frac{1}{36} \left( -7 P^2 Q_2 + 6 P_2^2 \right), \qquad Q_4 = \frac{1}{12} \left( -63 Q^2 (P_1^2 + P2) + 2 Q_2^2 \right),
\nonumber \\
&& P_5 = \frac{1}{60} \left( -7 P^2 Q_3 + 12 (P_1 P_4 + P_2 P_3) \right),
\nonumber \\
&& Q_5 = \frac{1}{20} \left( -21 Q^2 (P_1^2 + 6 P_1 P-2 + 3 P_3) + 4 (Q_2 Q_3 - Q_1 Q_4) \right),
\nonumber \\
&& P_6 = \frac{1}{90} \left( -7 P^2 Q_4 + 9 P_3^3 + 6 (P_2 P_4 - 10 P_1 P_5) \right), \qquad Q_6 = \frac{12 \left( P_4^2 + P_3 P_5 - 2 P_2 P_6 \right)}{7 P^2},
\nonumber \\
&& Q_7 = \frac{24 P_4 P_5}{7 P^2}, \quad Q_8 = \frac{3 \left(5 P_5^2 + 6 P_4 P_6 \right)}{7 P^2}, \quad Q_9 = \frac{30 P_5 P_6}{7 P^2}, \quad Q_{10} = \frac{18 P_6^2}{7 P^2}.
\end{eqnarray}
To find the relation between $P,\, Q$ and $P_1,\, Q_1$,  one should notice that the system of algebraic equations is over-determined, so there are additional relations between the coefficients.  Using the no-force condition~(\ref{chargemassrelation}) one finds from  the relations~(\ref{G2coeffs}) the following:
\begin{equation}
P_6 = \frac{343}{3240} P^2 Q^4
\end{equation}
In terms of $P_1, Q_1$ the no-force condition~(\ref{chargemassrelation}) reads
\begin{equation}
    \qquad 7 (P^2 + Q^2) = 3 P_1^2 + Q_1^2 - 3 P_1 Q_1,
\end{equation}
Solving these equations one has to take the solution corresponding to the positive values of $P^2,\, Q^2$ and positive values of $P_1, Q_1$. Lacking an analytic relation between the charges and $P_1,\, Q_1$, we cannot present the near-horizon expansion of the dilaton field, but we checked numerically that $\phi_{{\rm ext}} = \phi_h + O(r^5)$ and $\e^{2B} \propto r^2 + O(r^3)$ indeed.

The entropy of the extreme solution is
\begin{equation}
S_{{\rm ext}} = \frac{7 \pi}{6 (3^{2/7}) (5^{5/14})} \lvert P \rvert ^{9/7} \lvert Q \rvert ^{5/7}.
\end{equation}
Evaluation of mass confirms that $G_2$ black holes also have negative binding energy in the extreme case  (see Fig.~\ref{massesExtreme}).
The product of entropy of inner horizon $S_{-}$ and entropy of outer horizon $S_{+}$ obeys the following relation
\begin{equation}
S_{-} S_{+} = \pi^2 r_+^4 \frac{p_1^{9/7} q_1^{5/7}}{\mathcal{H}_1(1)^{3/7} \mathcal{H}_2(1)^{1/7}} = S_{{\rm ext}}^2.
\end{equation}

\subsection{Arbitrary $D$} \label{arbitraryDsection}
In arbitrary dimensions we have the solutions in the form~(\ref{intervalgeneral}). The functions $H_i$ obey the equations of motion~(\ref{EqHi}). Assuming the representation~(\ref{Hrepresentation}), we find that all the solutions obtained for $D = 4$ satisfy the equations of motion for an arbitrary $D$ if we replace combinations of $\kappa_j g_j^2$ in all the expressions for $P_i,\, Q_i$ and $g_i$ themselves by $\kappa_j g_j^2 / (D-3)^2$. The solutions have the same properties as their analogues in $D = 4$.

\section{Discussion}
We have constructed new analytic solutions of dilaton gravity with different couplings of the dilaton to electric/magnetic sectors. They correspond to $n = 3,\, 5$ sequence of triangular coupling and are presented as $B_2$ and $G_2$ Toda solutions respectively. The metric functions are expressed in terms of polynomials whose rank depends on the group and which were found explicitly for the non-extreme and extremal cases. As expected, the dilaton function has $n$ zeroes between the horizons in the non-extreme case while in the extremal case the integer $n$ is the power of the first Taylor expansion term of the dilaton at the degenerate horizon. The binding energy depends on the ratio of electric/magnetic charges and it becomes negative for all charges in the extremal case. New dyons satisfy the entropy product rule, indicating on possibility of statistical explanation of their entropy.

It would be interesting to find the higher-dimensional origin of the presented solutions. Clearly, the underlying group structure may serve the key. Indeed, it is believed that the Toda solutions are related to one-dimensional subspaces of the three-dimensional sigma models target spaces arising in dimensional reduction of higher-dimensional theories~\cite{Chemissany:2010ay, Chemissany:2010zp, Fre:2011uy}. This is so for $a = b$~\cite{Galtsov:1995mb}. The simplest sigma-model underlying to $B_2 = SO(2,3)$ could be the four-dimensional Einstein-Maxwell-Dilaton-Theory~\cite{Galtsov:1994sjr}, while the $G_2$ is encountered in the $D = 5$ minimal supergravity~\cite{Bouchareb:2007ax}. These theories, however, do not seem to be relevant. Rather, our Toda black holes could correspond to some higher-dimensional oxidation of the three-dimensional cosets~\cite{Cremmer:1999du, Clement:2013fc}, in particular, different embeddings of  $G_2$ are known~\cite{Galtsov:2008jjb}. We leave this question to future work.

\section*{Acknowledgements}
We thank G\'erard Cl\'ement and Eugene Davydov for reading the paper and useful comments. The work of CMC was   supported by the Ministry of Science and Technology of the R.O.C. under the grant MOST 106-2112-M-008-010. DG acknowledges the support of the Russian Foundation of Fundamental Research under the project 17-02-01299a and the Russian Government Program of Competitive Growth of the Kazan Federal University.

\end{document}